# Solute hydrogen and deuterium observed at the near atomic scale in high-strength steel

Andrew J. Breen[1,2,3*], Leigh T. Stephenson[1], Binhan Sun[1], Yujiao Li[4], Olga Kasian[1,5], Dierk Raabe[1], Michael Herbig[1], Baptiste Gault[1,5*]

[1] Max-Planck-Institut für Eisenforschung, Max-Planck-Straße 1, Düsseldorf, Germany

[2] School of Aerospace, Mechanical & Mechatronic Engineering, The University of Sydney, Sydney, NSW, Australia

[3] Australian Centre for Microscopy & Microanalysis, The University of Sydney, Sydney, NSW, Australia

[4] Zentrum für Grenzflächendominierte Höchstleistungswerkstoffe (ZGH), Ruhr-Universität Bochum, Germany

[5] Helmholtz-Zentrum Berlin, Helmholtz-Institute Erlangen-Nürnberg, 14109 Berlin, Germany

[6] Department of Materials, Imperial College London, Royal School of Mines, Exhibition Road,

[*] Corresponding author: andrew.breen@sydney.edu.au, b.gault@mpie.de

## Abstract

Observing solute hydrogen (H) in matter is a formidable challenge, yet, enabling quantitative imaging of H at the atomic-scale is critical to understand its deleterious influence on the mechanical strength of many metallic alloys that has resulted in many catastrophic failures of engineering parts and structures. Here, we report on the APT analysis of hydrogen (H) and deuterium (D) within the nanostructure of an ultra-high strength steel with high resistance to hydrogen embrittlement. Cold drawn, severely deformed pearlitic steel wires (Fe-0.98C-0.31Mn-0.20Si-0.20Cr-0.01Cu-0.006P-0.007S wt.%, $\varepsilon = 3.1$) contains cementite decomposed during the pre-deformation of the alloy and ferrite. We find H and D within the decomposed cementite, and at some interfaces with the surrounding ferrite. To ascertain the origin of the H/D signal obtained in APT, we explored a series of experimental workflows including cryogenic specimen preparation and cryogenic-vacuum transfer from the preparation into a state-of-the-art atom probe. Our study points to the critical role of the preparation, i.e. the possible saturation of H-trapping sites during electrochemical polishing, how these can be alleviated by the use of an outgassing treatment, cryogenic preparation and transfer prior to charging. Accommodation of large amounts of H in the under-stoichiometric carbide likely explains the resistance of pearlite against hydrogen embrittlement.

## 1 Introduction

H is the lightest and smallest of all the elements. It is also ubiquitous. Its presence within the structure of metals cannot be avoided, partly due to its high mobility. The interactions between H and metals play an essential role in the structure-property relationships of engineering alloys [1]. Hydrogen's most striking effect on materials is a sudden and often unpredictable decrease in ductility, toughness, and generally in the material's resistance to fatigue-crack propagation known as hydrogen embrittlement. This phenomenon was discovered as early as 1875 by Johnson but is still not fully understood to date [2]. A real-life example: the oil spillage in the Gulf of Mexico in 2012 is attributed to the catastrophic failure of connector bolts due to hydrogen embrittlement[3]. Hydrogen embrittlement mostly affects structural, high-strength materials that find application in the e.g. automotive, aeronautical, oil and gas and power generation industries.

High-strength steels are amongst those materials that are particularly affected by hydrogen embrittlement. As reviewed recently by Bhadeshia [4], limited strategies have proven efficient to prevent such embrittlement. Among high-strength steels, pearlite exhibits both exceptional strength [5] and a high resistance to hydrogen embrittlement (HE) compared to other steels [6]. Pearlite consists of alternating layers of cementite and ferrite, formed through a diffusive eutectoid decomposition reaction of the austenite phase. Cementite has the stoichiometry $Fe_3C$ and its crystal structure is orthorhombic. Ferrite, has body-centered cubic structure and is Fe-rich. Upon severe wire drawing deformation, the cementite decomposes, driven by dislocations penetrating through it, leading to a reduction in the carbon content in the cementite and hence leaving carbon-vacancies within the structure [7]. Trapping diffusive H at specific microstructural features (interfaces, precipitates) to prevent its accumulation and thereby mitigate its influence on the properties is a possible strategy.

Direct atomic-scale observation of H is limited, yet, of fundamental importance for understanding the mechanisms governing its detrimental effects in metals and the resultant influence on the macroscopic properties. Atom probe tomography (APT) is the only microscopy and microanalysis technique to allow for direct observation of H in three-dimensions at the near atomic-scale, with high potential for possible quantitative analysis of H. APT is poised to play a key role in the investigation of H in materials. APT exploits an intense electric field to progressively erode the specimen, almost atom-by-atom. Upon processing of the data, APT provides 3D elemental mapping with near-atomic resolution in volumes routinely of 100 x 100 x 1000 $nm^3$ [8]. Steels were one of the prime targets for H-studies by APT, in particular seeking to reveal the trapping behaviour of H/D at carbides [9–11].

APT has the intrinsic capacity to detect H, however, although the analysis chamber is kept at an ultra-high vacuum, residual H is detected during every experiment. This H likely originates from the chamber wall made of stainless steel in most instruments, as well as from adsorbed species on the surface of the specimen or the specimen holder. It is however challenging to determine whether H ions originate from the specimen or from ionization of the residual gas from the analysis chamber. This severely hinders the potential for deploying APT to investigate H in materials. One approach proposed to mitigate these issues is to use isotopic marking. The specimen can be charged with deuterium (D or $^2H$), which has a very low natural isotopic abundancy and has been used in several studies [9,12–14].

Charging can be performed through either electrochemical or gas-phase charging. For deuteration to be successful, there must also be microstructural trapping sites available and the specimen must not have been saturated with H. The high diffusivity of H/D in most materials means that the analysis must be performed quickly following charging otherwise, the D desorbs from the specimen, as successfully demonstrated by Haley et al. [14]. If the specimen may be cryogenically cooled immediately after charging, which slows down outward diffusion, the likelihood of successful detection of D increases [9–11]. The distribution of D can then be studied to reveal possible locations of segregation inside a material. With recent developments in cryogenic transfer dedicated to APT[15,16], the potential for success has increased significantly. In addition, forays into cryogenic specimen preparation by focused-ion beam (FIB) have also been reported[17], demonstrating how important this is to prevent the pick-up of H during FIB-preparation [18]. Yet much remains to be done to make the analysis of H in metals routine, and quantitative.

Results on H/D analysis using APT for solutes and stable hydrides remain limited but reports can be found in the literature for multilayered materials[13,19], Ti-based [20–23] or Zr-based alloys [24] for instance. Deuteration is however not without challenges since H can be detected both as atomic ions, i.e. $H^+$, as well as molecular ions, i.e. $H_2^+$ and $H_3^+$. The relative amount of these different ions depends strongly on the strength of the electrostatic field during the APT analysis [25] making quantification particularly complex [26].

Here, we investigate the distribution of H in a pearlitic steel produced by severe cold drawing of a wire with hypereutectoid composition. We report on a uniquely thorough set of analyses of specimens prepared by electrochemical polishing and focused-ion beam (FIB) milling at room and cryogenic temperatures, with and without cryogenic vacuum transfer, with and without D-charging. Yet, this article is not about a mere methodology study since reliably detecting and discussing H/D and their traps in complex materials requires to operate on the basis of such a wide probing matrix, including many preparation and charging conditions, as otherwise the H/D signal is too unreliable.

Indeed, we demonstrate that H in all APT datasets originates mostly from specimen preparation when using electrochemical polishing as well as the FIB. D-charging of specimens saturated with H from the preparation is inefficient. Our study sheds light onto the origins of the contamination by H, also that the H-distribution revealed by APT can likely be used to highlight H-trapping sites within the bulk of a material, yet precise quantification still likely requires isotopic marking and precise physics-informed statistical analyses. From a materials standpoint, both H and D are found segregated to the partly-decomposed cementite, with a slight tendency to segregate at some interfaces. Our result rationalise the importance of the defects associated to the off-stoichiometric composition of the decomposed cementite as trapping sites for hydrogen, which help prevent hydrogen embrittlement.

# 2 Materials and methods

## 2.1 Materials

The material used in this work was a severely deformed cold drawn pearlitic steel wire with hypereutectoid composition (Fe-0.98 C-0.31Mn-0.20Si-0.20Cr-0.01Cu0.006 P-0.007 S, wt.%). The initial diameter of the wire was 0.54 mm with an interlamellar spacing of 67 nm. The wire was then cold drawn to a true (logarithmic) strain of 3.10. A tensile test was then performed on the wire at room temperature and at a constant strain rate of $2 \times 10^{-3} s^{-1}$. The tensile strength of the wire was measured to be 3.85 ± 0.02 GPa. Further analyses were performed on the as-cold-drawn material. The performance of this wire, and a range of other wires with varying cold drawn strain values, were previously reported in [7]. This wire was specifically chosen because it was likely to contain numerous decomposed cementite and ferrite lamellae within one atom probe dataset, enabling advantageous conditions to study the H trapping sites around these interfaces and within the different phases. The wire also had a favourable geometry for specimen preparation by electrochemical polishing.

## 2.2 Specimen preparation and imaging

Needle-shaped specimens for APT were prepared, post-tensile test, by electrochemical polishing using a two-stage process: first rough polishing was performed in a solution of 25% perchloric acid (70%) in glacial acetic acid with 10 – 20 V until the specimen's end was roughly below a micron in thickness; second, a solution of 5% perchloric in butoxyethanol with 10 – 20 V was used to finalize the shaping of the specimens. A dual beam scanning electron microscope / Xe-plasma focused ion beam, FEI Helios PFIB, was used to sharpen electrochemically polished blanks with an end-size in the range of 20 µm. For transmission electron microscopy (TEM), specimens were prepared by in-situ lift-out from the cross-section of the wire, using a dual beam FEI Helios 600. The inverse pole figure (IPF) and image quality (IQ) maps were generated by nanobeam diffraction orientation mapping in a JEOL JEM-2200FS TEM operated at 200 kV using the commercial setup ASTAR by Nanomegas at 0.5 nm spot size, 1 nm step size, 20 ms exposure times and without precession. An IPF map of only the ferritic phase was generated using the commercial indexing software by Nanomegas. All points above 150 indexing quality and 5 reliability were exported and further processed in the TSL 7.0.1. software by EDAX.

## 2.3 Specimen charging

Specimen charging and transfers were performed using the facilities from the Laplace project at the Max-Planck-Institut für Eisenforschung[16]. Specimens were charged with deuterium using an electrochemical cell and 0.1 M NaOD in $D_2O$ charging solution introduced in ref.[14]. For simplicity, ease of handling and more rapid transfer, some modifications to this previously reported setup were made. The Nafion® membrane around the sample was removed to simplify handling. A new sample holder was constructed at the working electrode to hold an APT multiwire puck and 4 electropolished needles (see Figure 1a). Immediately after charging in one of the explored workflows (see section 3.2.), the puck was removed from the charging cell and inserted into an Al puck holder block (Figure 1b). Figure 1c shows the cell configuration and potentiostat attachment. For all charging experiments, a PalmSens® EmStat3® potentiostat was used. Figure 2c shows the charging cell and potentiostat connection assembly.

Complementary testing of the charging setup was performed using a cast ultra-fine-grained, fully pearlitic, high carbon steel of composition Fe-0.74C (w.%). The cold drawn wire was not used for these experiments primarily because the geometry was not suitable for thermal desorption spectroscopy (TDS). Unlike the pearlitic wire, this material was not exposed to plastic deformation. However, the results still provide valuable information of how a fine pearlitic microstructure behaves under the charging conditions used.

Figure 1d shows the resulting cyclic voltammetry curve for a test blank from the Fe-0.743C sample submerged approximately 5 mm into the charging solution with a cross section of 0.5 x 0.5 mm. The voltage was swept from -3 to +3 V. The hydrogen evolution reaction (HER) and the oxygen evolution reaction (OER) regions can clearly be observed. A constant voltage of -1.2 V was found to generate the HER with measurable current response while suppressing bubble formation. The same charging solution and applied voltage had been previously reported by Haley et al. [14] in a similar setup for charging high-Mn steel.

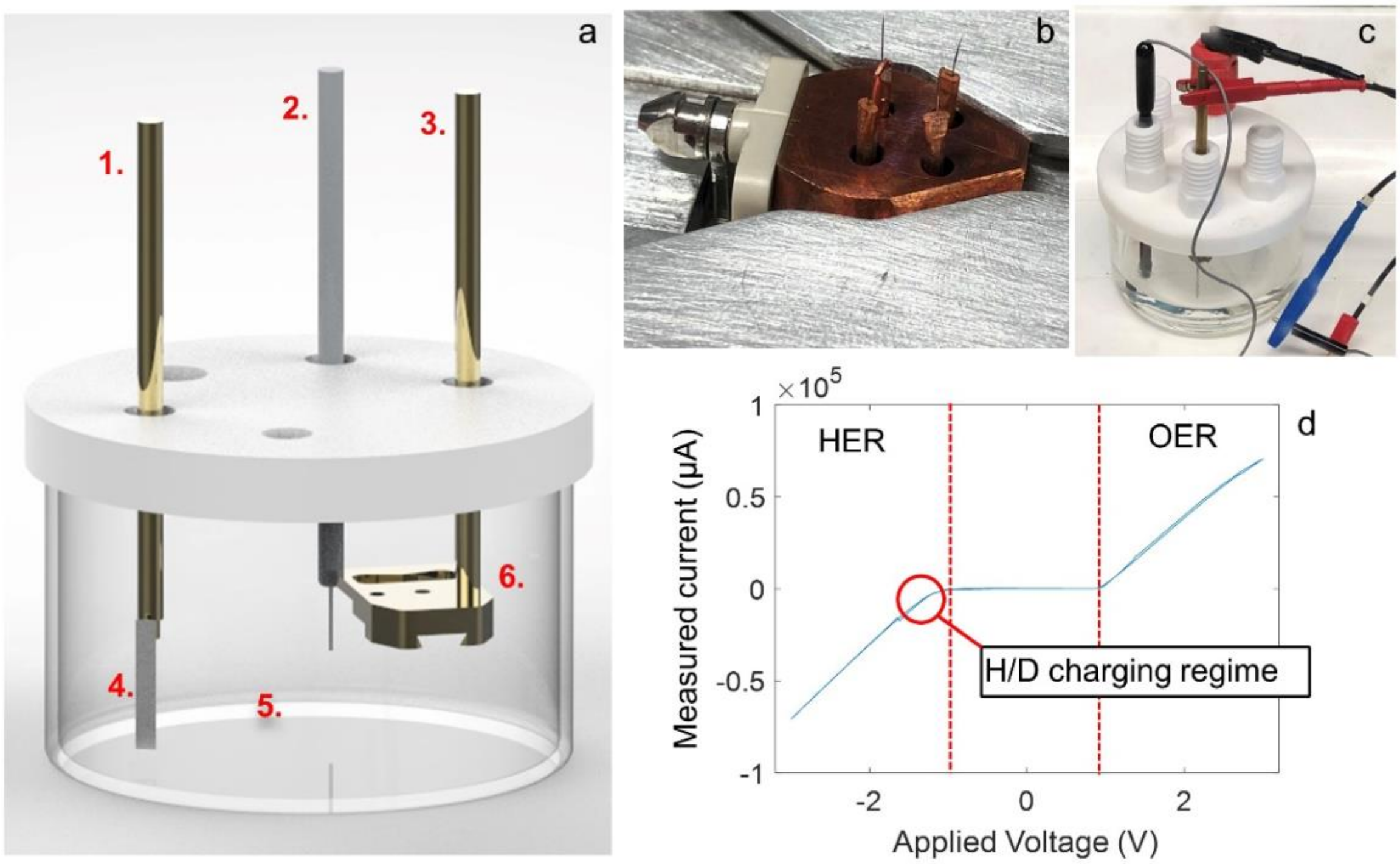

*Figure 1: Electrochemical H/D charging setup. (a) Schematic of the cell. 1. The Counter electrode (+). 2. The reference electrode (Ag/AgCl). 3. The working electrode (-). 4. A Pt wire or foil (which has a surface area larger than the specimen/s being charged. 5. Glass Beaker which is filled with 0.1M NaOD in $D_2O$ so that only the specimens and electrodes are in contact. 6. Special puck holder attachment (so that specimens can be charged directly in puck). (b) Wire specimens in cryo multi-wire puck in Al block which could be precooled to LN2 temperatures. (c) testing of electrochemical cell with pearlite blank. (d) Cyclic voltammetry curve showing current response from pearlite test blank. -1.2 V was found to be a good working voltage for a H/D charging regime.*

The possibility of electrochemically charging this pearlitic microstructure with deuterium was confirmed using TDS. Two 14 x 14 x 0.5–1 mm plates of the Fe-0.743C sample were mechanically polished. A scanning electron microscope (SEM) image taken in a Zeiss Merlin at 3 kV acceleration voltage using the inLens detector is shown in Figure 2a and confirms the fine pearlitic microstructure similar to that of the wires considered here.

One plate was then directly analysed for H/D content using TDS while the other was charged with D for 90 mins, followed by TDS analysis. Care was taken not to submerge the supporting alligator clip into the charging solution so that the H/D evolution reaction would occur exclusively on the TDS sample not at the support. This meant that only approximately 50 % of the TDS sample was submerged in the solution. The time interval between the end of D charging and the start of TDS experiments was less than 20 min. The TDS was performed on a home-built ultra-high vacuum-based setup[27] at a constant heating rate of 26°C·min$^{-1}$. The gases evolving during the continuous heating of a sample were detected by an MKS MicroVision 2 quadrupole mass spectrometer. The resulting spectra are displayed in Figure 2b, showing a clear indication of a peak associated to the charging of H inside the specimen at approx. 500°C.

The TDS analysis was performed only to indicate whether our electrochemical charging process was effective. The position of peaks in TDS is highly dependent on the heating rates. Higher heating rates shift the TDS peaks to higher temperatures. Here, we uses a rather rapid heating rate (26°C/min) compared with more conventional, commercial TDS setups which are normally operated at around 1~5°C/min. This TDS provides qualitative. Indeed, a full Kissinger analysis cannot be performed here, since if would have required a thorough exploration of heating rates to determine the trapping energy associated to the peaks. We still obtain a higher cumulative amount of H/D upon charging, as readily visible on the plot. In addition, it took approx. 20min between the end of charging and the start of TDS, which might be already enough to lose some H from the most shallow traps.

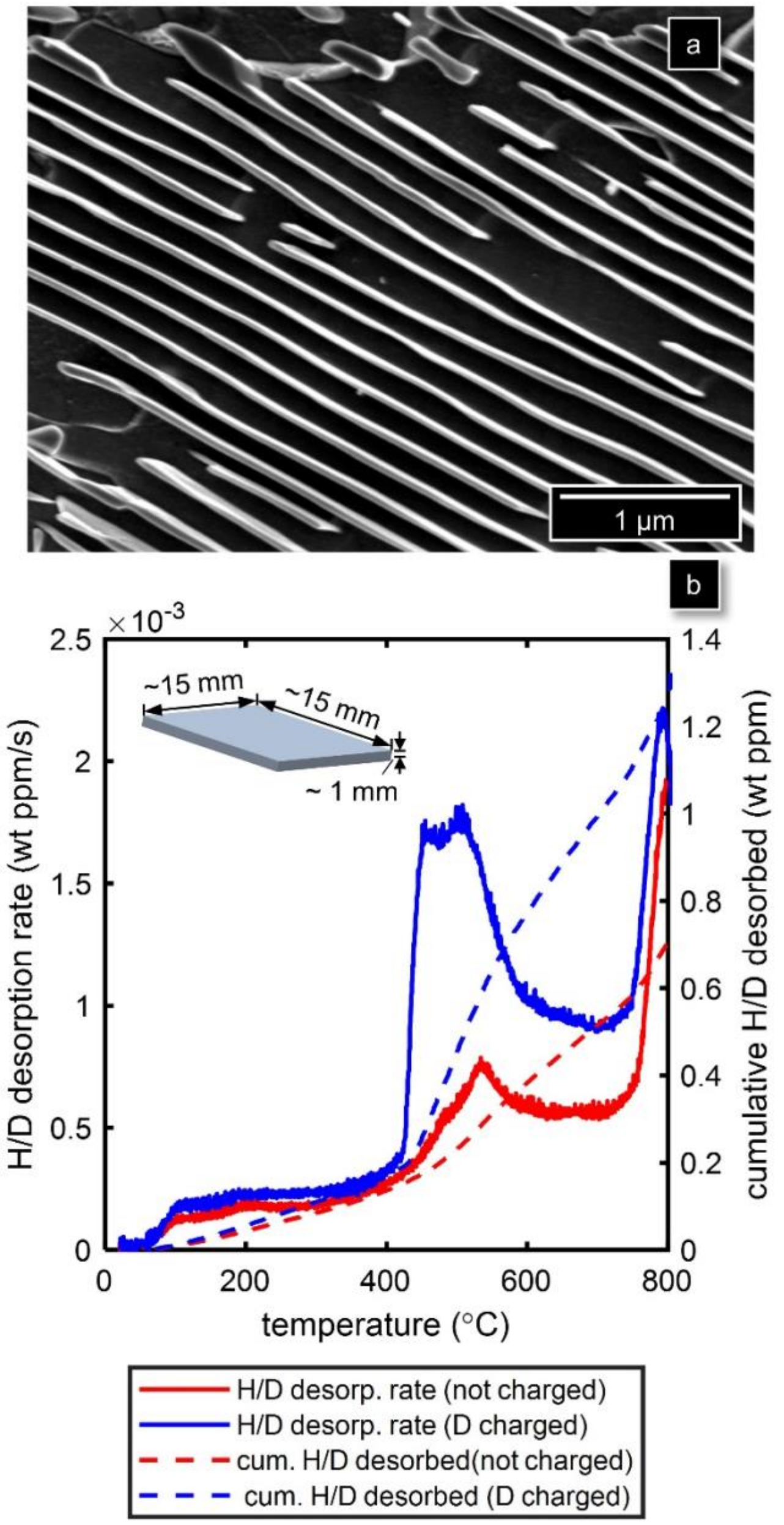

*Figure 2: (a) A SEM image of the fine pearlite microstructure of the cast Fe-0.743C sample. (b) The TDS curve showing significantly higher D detection after charging, with the desorption rate on the left axis and the cumulative amount of H/D on the right axis.*

## 2.4 Specimen cryo-transfer

In workflow W5, after charging, the block with puck and specimens was initially plunged into LN2 but the much higher temperature of the puck and samples, vigorous bubbling, along with thermal shock, was at risk of damaging the specimens. Instead, the puck and specimens were only passively cooled on the pre-LN2 cooled Al block. This was done in air and consequently ice formed on the specimens. However, after loading the samples into the buffer chamber of the PFIB docking station, this ice quickly sublimated under the high vacuum conditions. The samples were then transferred directly into the Ferrovac® suitcase. Within the suitcase, samples could be kept at temperatures of approximately –180 °C with a vacuum of $10^{-8}$ - $10^{-7}$ mbar and transferred directly to the buffer chamber of the LEAP 5000 XS/XR instruments.

We did not succeed despite several attempts at charging and cryogenic cooling carried out inside a glovebox filled with $N_2$ and a highly controlled level of moisture, followed by direct transfer into the atom probe via the cryo-vacuum transfer suitcase. Specimen yield was low, despite starting to field evaporate at low voltages, suggesting the specimens had successfully survived charging and transfer. Strong peaks in the mass spectrum pertaining to Cu and W peaks could be observed, which suggested contaminants from previous charging attempts frozen onto the surface of the needles. Freshly prepared charging solution and more careful cleaning of the specimens with isopropanol before cryogenic cooling is highly necessary.

### 2.5 Atom probe experiments

Most of the APT data were acquired on a Cameca LEAP 5000 XR with a base temperature of 60 K, in high-voltage pulsing mode with a pulse fraction of 15-20% and at a repetition rate of 200-250 kHz. Similar parameters were reported to yield accurate compositions in the analysis of these materials[5]. A detailed list of the experimental parameters for each representative dataset of the workflows is provided in the supplementary information. Voltage pulsing was chosen primarily to minimize the formation of $H_2^{+}$ ions which complicate the quantification of deuterium which was to be charged into the specimens. Voltage pulsing also had the added benefit of improved spatial resolution which could be used to facilitate crystallographic indexing of the datasets as shown in Figure 3. APT parameters were intentionally kept approximately constant so that changes to the H and D measurements could be more confidently attributed to the different experimental workflows and not experimental running conditions. Nevertheless, some data was also collected on a Cameca LEAP 5000 XS and LEAP 3000 and using laser pulsing so the influence of these changes to APT running conditions could be assessed. Data reconstruction and processing was performed in Cameca IVAS® 3.8.2 and a range of specifically coded routines in Matlab.

## 3 Results & Discussion

### *3.1* Material crystallographic characterization

Nano-beam diffraction (NBD) and bright field scanning transmission electron microscopy (BF-STEM) were performed on a transverse cross-section of a wire drawn to a true strain of $\varepsilon = 3.10$ for information on crystallographic texture. The same specimen was then analysed by APT. Figure 3(a) is the inverse pole figure (IPF) map ($Z$) – where $Z$ is the direction normal to the imaging surface and corresponds to the drawing direction of the wire (only ferrite was indexed since indexing cementite at these sizes and deformation states was not possible). Figure 3(b) is the corresponding image quality (IQ) map on which the grain boundary misorientation angle map was superimposed. Figure 3(c) is an APT reconstruction of the wire (only Fe and C atoms are shown for clarity). APT performed on this material in voltage pulsing mode yields high quality data in which partial crystallographic information can also be extracted. Although crystallographic poles were difficult to distinguish from the density variations, due to the decomposed cementite lathes, regions containing clear lattice planes were observed (Figure 3(d)). Spatial distribution maps[28,29] (SDMs) were calculated in the depth of a subset of the reconstructed data in the pole regions, confirming the presence of (110) planes within the reconstruction, as shown in Figure 3 (e). This information was used for guiding reconstruction calibration wherever possible by adjusting the reconstruction parameters to obtain appropriate plane spacings (measured from the peak-to-peak distances). These snapshots of the ultrafine and highly complex microstructure of the pearlitic wire highlight the need for high-resolution correlative structural and chemical characterization tools, as well as the inherent high quality of the collected APT data. Only the use of such delicate probing protocols can provide clues as to the distribution of H and D and the associated trapping behaviour, and, potentially the influence of the crystallographic character of the interfacial regions.

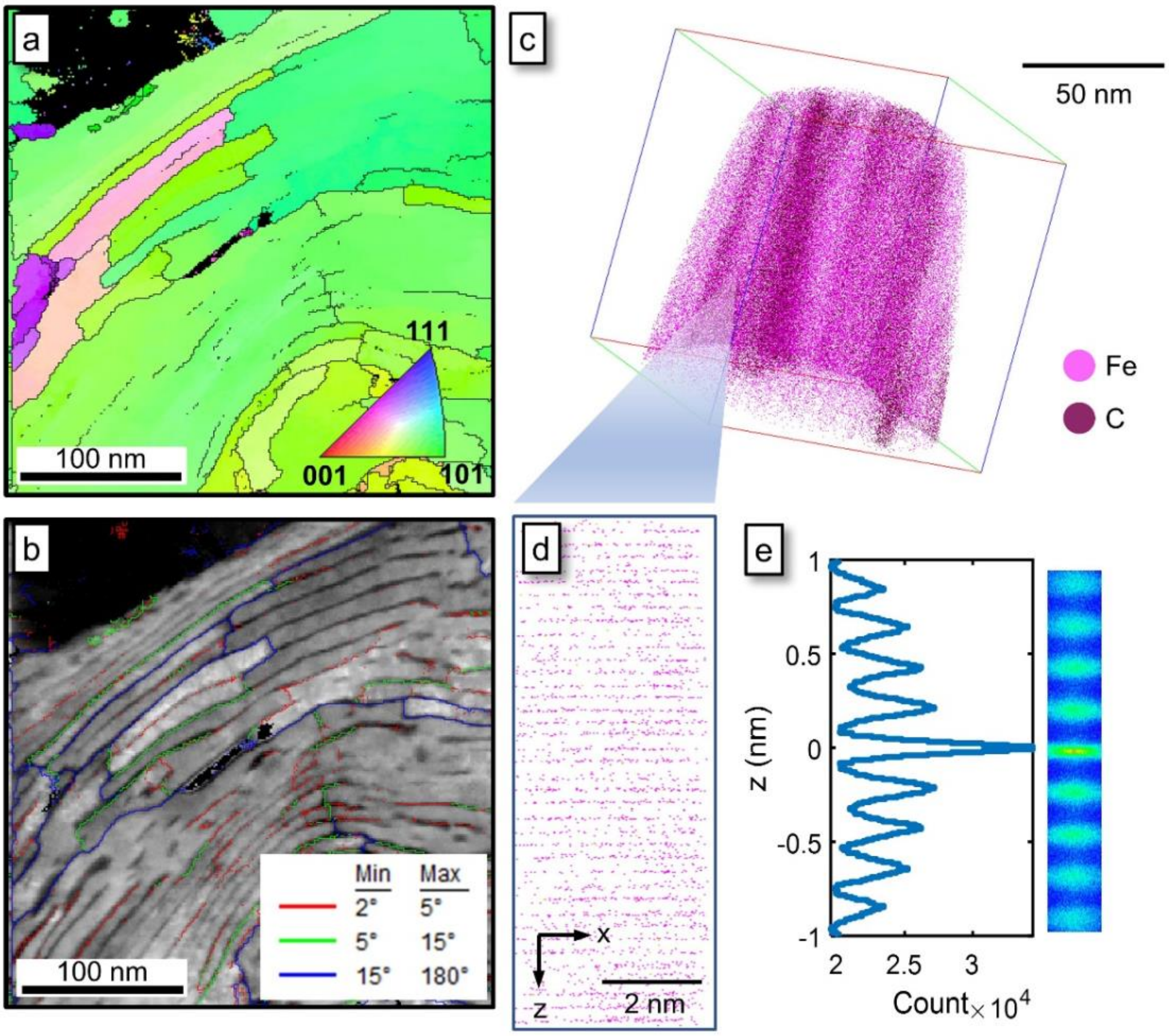

*Figure 3:microstructural analysis of the wire: (a) Inverse pole figure – Z from the nanobeam diffraction analysis of the ε = 3.10 deformed wire showing microstructure and texture; b corresponding bright field scanning transmission electron micrograph, on which the grain boundary misorientation angle map was superimposed. (c) representative APT reconstruction of the wire (only Fe and C atoms are shown for clarity. (d) a zoomed in region of the reconstruction showing individual lattice planes. (e) SDM of region in d showing clear lattice periodicity.*

## 3.2 Experimental workflows for H and Deuterium charging and probing

It is essential to develop and test protocols that can establish routine acquisition of high quality, reliable and reproducible data in order to inform atomic-scale H detection with well-determined, robust estimations of the measurement errors. Seven different workflows were followed between electropolished specimen preparation to APT analysis, as laid out in Figure 4 and detailed below:

1. The specimen was directly inserted into the load-lock of the atom probe and ran in voltage pulsing mode.
2. The specimen was directly inserted into the load-lock of the atom probe and ran in laser pulsing mode.
3. The specimen was placed into the load-lock of the atom probe and heat treated under high vacuum conditions (below $10^{-7}$mbar) for up to 4h at 150°C and ran in voltage pulsing mode.
4. Following this heat treatment for 4h at 150°C, the specimen was removed from the load-lock and electrochemically charged with D and transferred into the atom probe via the airlock, at room temperature, and ran in voltage pulsing mode.
5. Following heat treatment, the specimen was electrochemically charged with D, passively cooled on an $LN_2$ cooled block and transferred into the atom probe via a UHV cryogenically cooled suitcase and ran in voltage pulsing mode.
6. Following rough electropolishing down to a size in the range of 2–10 μm, the specimen was sharpened further at cryogenic temperatures (~ -130°C) in the PFIB before being transferred into the PFIB/UHV buffer chamber and then into the UHV cryogenically cooled suitcase. The details of the set-up and conditions are described in [16].
7. Following rough electropolishing down to a size in the range of 2–10 μm, the specimen was sharpened further in the PFIB at room temperature before being transferred into the atom probe in air. Samples were ran using voltage pulsing.

The experimental workflows are non-exhaustive, other variations were attempted, but these 7 represent the workflows for which sufficient APT data was obtained and presented here. Workflow 1 (W1) was designed as the basic 'control' experiment to which the other workflows were compared.

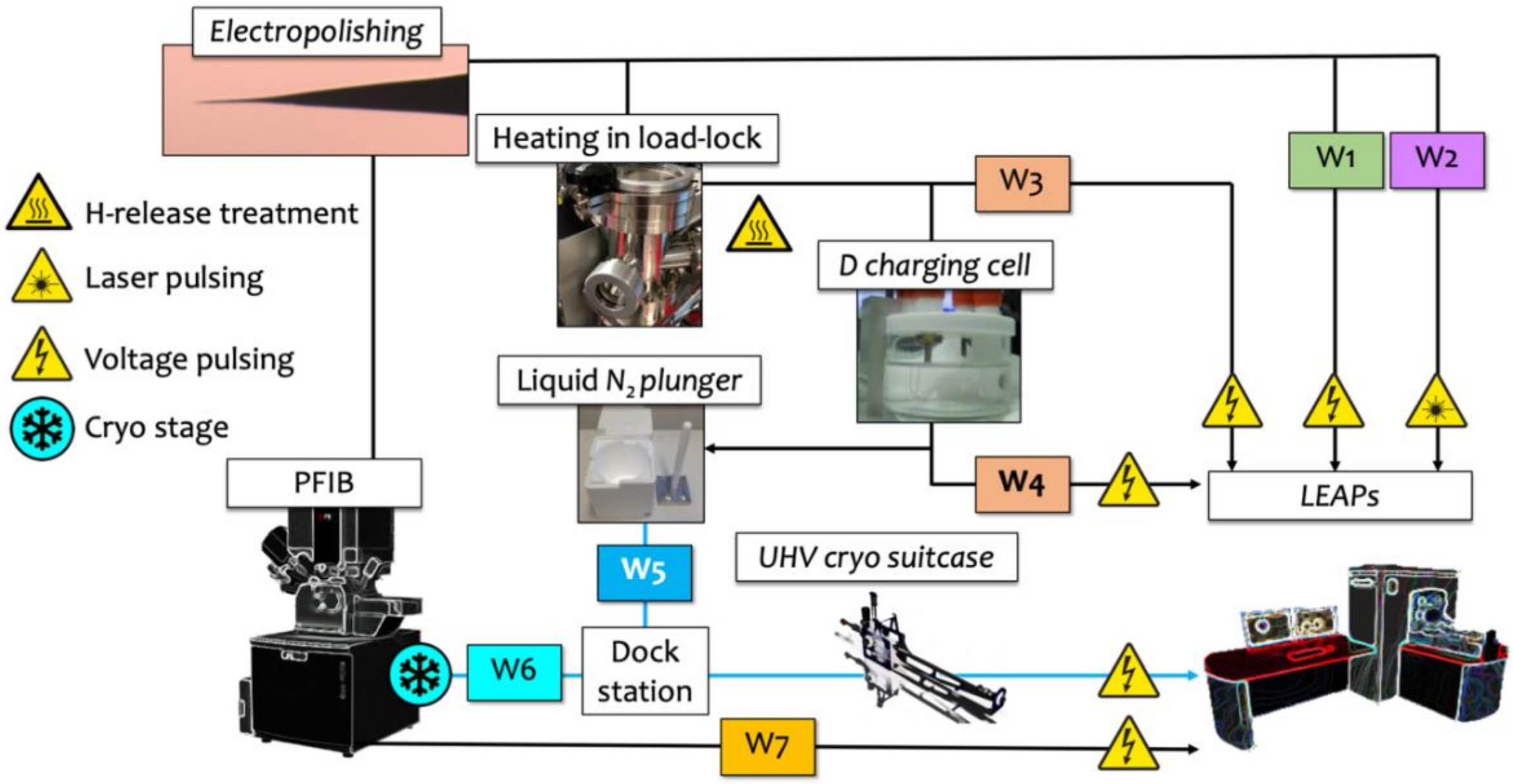

*Figure 4: Seven experimental workflows (W1 – W7) from an electropolished specimen to analysis in the atom probe. D charged workflow numbers have been bolded. Cryogenic transfer experiments are*

*coloured blue. PFIB: Xe-plasma Focussed Ion Beam Instrument. LEAP: LEAP Atom Probe Tomography Instruments.*

A summary of the APT results obtained from representative datasets from each of the 7 experimental workflows outlined previously are shown in Figure 5. Atom maps of the C and H/D are provided along with corresponding mass spectra (normalized by total atom counts) so that the change in H and D signal behaviour between the different datasets can be assessed. The bulk C, H, D concentrations for each workflow are also provided in Figure 6. In all datasets, H clearly partitions to the decomposed cementite lamellae, where C content is higher.

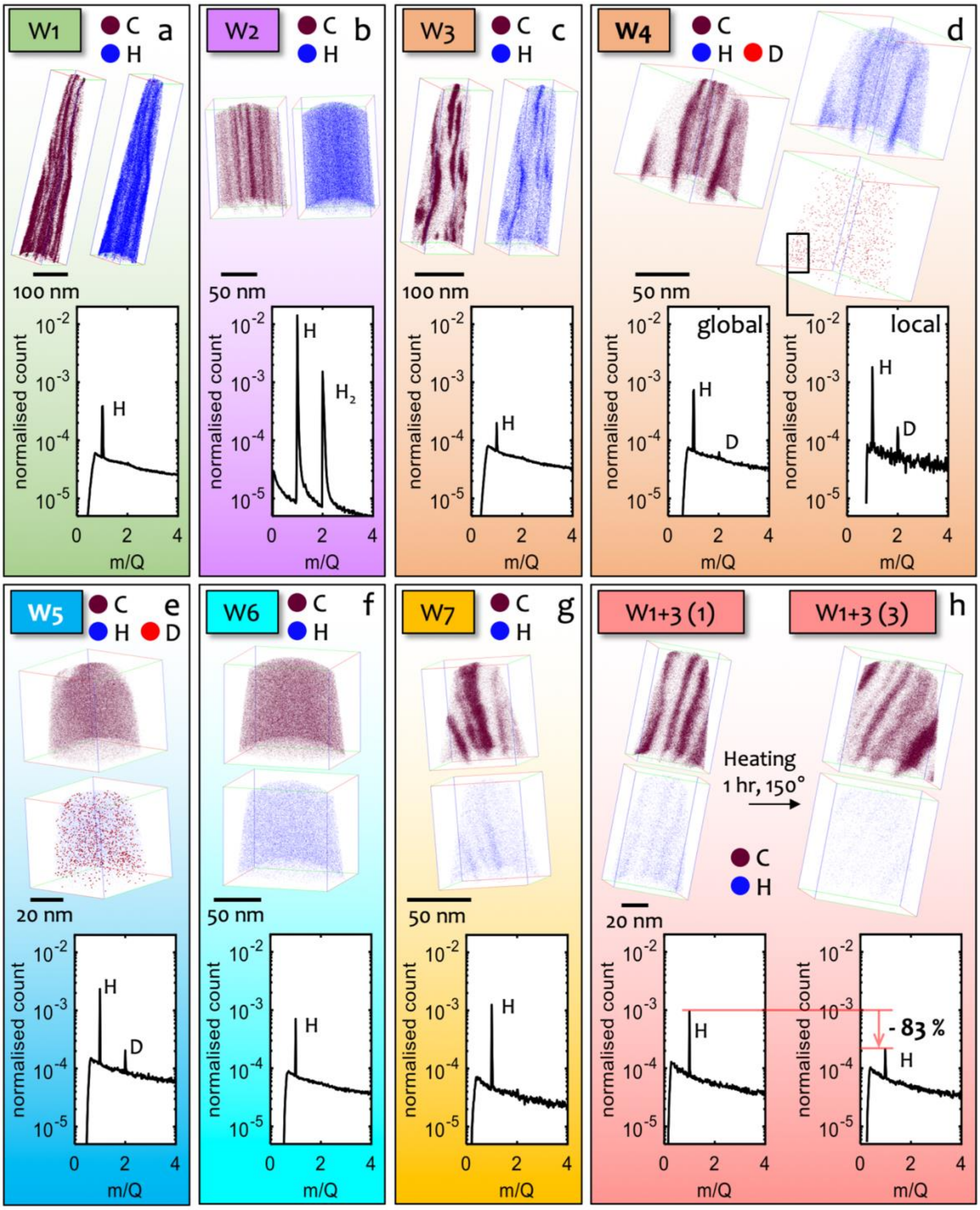

*Figure 5: APT results obtained from the deformed pearlitic wire after the different experimental workflows. (a) – (g) correspond to the results from workflows W1 – W7, respectively. (h) represents the results from the same specimen after electropolishing (workflow W1) and then after heating in the load-lock for 1 hour (workflow W3). The APT reconstructions (H and C maps for clarity) and the accompanying H peaks in the mass spectrum are shown for each workflow. The results not only reveal the strong link between specimen preparation, charging and specimen transport but also the robustness and quantification challenges associated with H detection.*

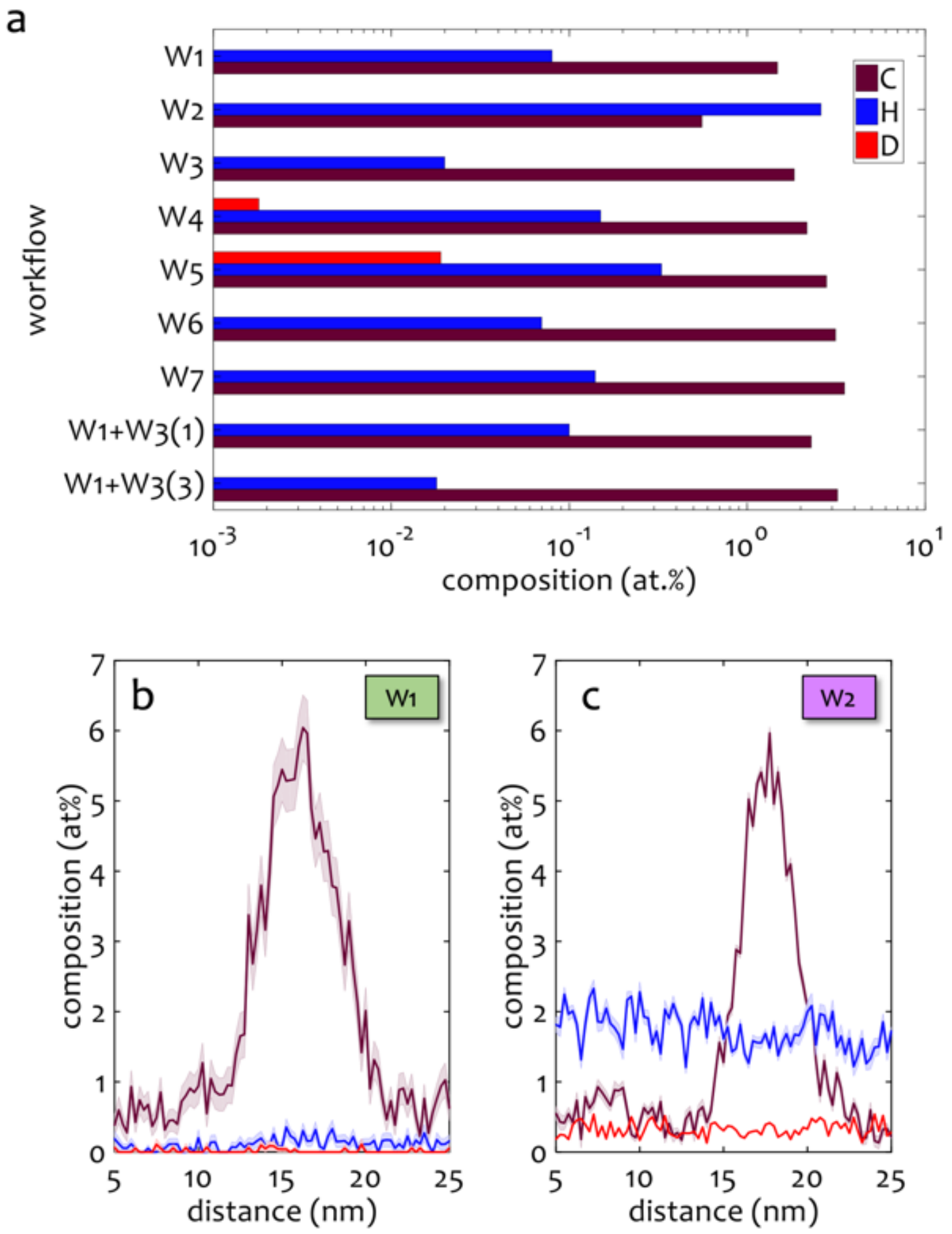

*Figure 6: (a) bulk C, H and D concentrations (at. %) for the respective workflows. Bulk calculations were background corrected on IVAS® 3.8.2. Composition profile through a decomposed cementite lath for data obtained through W1 in (b) and W2 in (c).*

## 3.3 High voltage vs. laser pulsing effects on H detection in atom probe tomography (W2)

In Figure 5a, which represents the 'control' type experiment whereby the specimen was electropolished and immediately analyzed using voltage pulsing, H exclusively evaporates as the atomic ion $H^+$, detected in the peak at m/q = 1 Da. This behavior was consistently observed for all datasets collected from workflow W1. However, when laser pulsing was used in workflow W2 (Figure 5b), a significant proportion of total H (32%) evaporated as a molecular ion ($H_2^+$). A much higher overall composition of H was also observed, i.e. 2.6% as compared to < 0.1% for all voltage pulsing measurements. This is evidenced by the difference in the base level of $H^+$ and $H_2^+$ in the composition profiles calculated through a cementite lath in datasets obtained using workflows W1 and W2 shown respectively in Figure 6b and 4c. In Figure 6c, the H-signal inside the lath even appears to drop, which can be explained by the local increase in the electrostatic field, as explain below).

Our results from W2 suggest that laser pulsing promoted the detection of spurious H. This H could originate from the field ionization of residual H from the vacuum chamber in its gaseous form. It is also likely that the detection of H comes, at least in part, from the field desorption of residual H that was before adsorbed on the specimen's surface. This H can come from the chamber, the preparation routine, and potentially migrated along the specimen's shank, driven by the gradient in electrostatic field. The higher electrostatic field conditions obtained in using voltage pulsing (W1) lead to better screening of the specimen's apex against contamination by H. These results confirm once again that voltage pulsing is better for quantifying H originating from within the specimen, in agreement with previous reports in the literature[11,14,23]. Voltage pulsing was therefore used in all remaining workflows despite the higher chance of specimen fracture and lower yield.

## 3.4 Effect of specimen pre-heating (W3) on H detection

In experimental workflow W3, the specimen was heated to ~150 ºC for approximately 4 hours in the load-lock of the atom probe. The heater used is the one typically used to clean the local electrodes inside the vacuum chamber of the atom probe, and the temperature cannot be adjusted. Figure 4c indicates that the relative H content declined relative to workflow W1. This is suggestive of outgassing of H contained within the specimen.

However, a certain uncertainty still remained as to whether the reduction in H was due to a difference in the microstructure between specimens. Indeed, the relative volume fraction of decomposed cementite and thickness of lamellae will influence bulk H detection[7]. We performed another experiment, whereby the same specimen was analysed under workflow W1 and then workflow W3 (Figure 5h). We confirmed that heating of the specimen, in this case for only 1 hour, was sufficient to reduce the H detected by 83% without appreciable carbon redistribution or change in the underlying microstructure. The results further add proof to the argument that the majority of H being detected under these experimental conditions is in-fact from the sample and not from the chamber, otherwise, such a drastic decline in H detection after heating would not have been expected.

## 3.5 Effect of D-charging (W4) & cryogenic transport (W5)

Charging conditions are a critical consideration for successful D detection. Using our home-built setup, we charged the specimens for 1 hour, which was sufficient to introduce D within the sample, as confirmed by complementary Thermal Desorption Spectroscopy (TDS) experiments. The conditions used, also did not appear to damage the specimens, although some solution residue on the tip surface was suspected due to lack of cleaning after polishing. Several experiments with charging times of only 5 minutes were conducted where no successful D was observed.

A small peak at m/q = 2 Da was observed in some of the datasets obtained following workflow W4, as shown in Figure 5d. The ions corresponding to this peak were seen to segregate non-uniformly to the cementite/ferrite interfacial regions. Since H was expected to only evaporate as a single ion under these experimental conditions, this signal likely originated from D located at strong trapping sites at these positions. Further investigation into the local electric field, shown later in the manuscript, also provides evidence for this peak being D and not simply $H_2$ brought about by a lower electric field. In workflow W5, after deuterium charging and cryogenic transfer, an even stronger peak at m/q = 2 was observed (0.75 at. %), which has been attributed to D. The higher carbon content (2.17 at. %) suggests that a large decomposed cementite region was captured in this dataset, or, considering the size, possibly a region of retained austenite, which is rather unexpected in these materials.

## 3.6 Cryogenic vs. room temperature focussed ion beam preparation (W6 & W7)

Figure 5f and g show the results from PFIB specimen sharpening at cryogenic temperature in W6 and room temperature in W7. Bulk H composition in the specimen sharpened at cryogenic temperature was only 0.07 at. % as opposed to 0.14 at. % when sharpened at room temperature, even though the cryogenic PFIB specimen captured a decomposed cementite region where H concentration would usually be expected to be higher. The H composition was also compared to the decomposed cementite region in W1. An iso-composition surface of C = 3.05 at. % was used to approximately isolate the decomposed cementite regions and the H content was found to be 0.21 at. % - significantly higher than what was observed in the cryogenically PFIB prepared specimen. The results suggest that cryogenic PFIB suppresses H uptake into the specimen. This supports the findings recently reported by Chang et al.[18], which focused on Ti-based alloys and further suggests this phenomenon holds true for ferrous alloys also, albeit to a lesser extent. Also interesting to note is that Figure 5e and f both capture larger decomposed cementite regions under the same APT experimental parameters, yet the peak at m/q = 2 Da is only seen in Figure 4e, further suggesting that this peak can be attributed to charged D and not $H_2$.

## 3.7 Influence of local electric field and carbon concentration on H measurements

Care must be taken in the interpretation of APT data when comparing relative H/D measurements between the workflows because of numerous factors which influence these measurements. Reports exist in the literature on the relationship between the strength of the electrostatic field and the detection of H either as atomic or molecular ions[25]. This also holds true for the carbon ions[30–32], in part because of the complex nature of the field evaporation and dissociation processes for molecular ions[33]. The influence of the local electric field on the carbon and H detection must hence be considered. We tracked these metrics and compare between datasets by voxelising[34] each dataset and estimate the local field, C and H contents in each voxel. Several different voxel sizes were trialled and cuboidal voxel sizes from $2^3$ $nm^3$ - $8^3$ $nm^3$ were found to provide a good combination of resolution and signal.

To measure the relative local field, the charge-state-ratio of the same atomic species of atom can be compared. The ratio of the peaks will vary with changes to electric field with higher charge states more likely to occur at higher fields, as predicted through post-ionization theory[35]. The clearest occurrence of multiple charge states for the same ion species across the datasets collected was for carbon. Peaks at m/q = 6 Da and m/q = 12 Da, containing ${}^{12}C^{2+}$ and ${}^{12}C^{1+}$ were observed. Some molecular C may have also been present in these peaks, but this was expected to be low, particularly in the voltage pulsed runs. Therefore, higher ${}^{12}C^{2+}/{}^{12}C^{1+}$ ratios can be attributed to relatively higher local electric fields.

The distribution of C composition vs. relative electric field in the representative dataset corresponding to workflow W1 is shown as an intensity histogram in Figure 7a. The ${}^{12}C^{2+}/{}^{12}C^{1+}$ ratio was calculated in each voxel and a 2D histogram of the results was plotted. Most voxels contain a very low carbon content of less than approximately 1 %, corresponding to the ferrite, where the electric field is relatively lower. The ${}^{12}C^{2+}/{}^{12}C^{1+}$ ratio increases sharply with only small increases in C content up to approximately 2 %, corresponding to the decomposed cementite. Any further increase in C does not appear to influence the local electric field significantly. No region was found to contain C compositions close to the stoichiometric carbon content of cementite of 25%. Similar trends were observed for all representative datasets and can be seen in the supplementary information.

The relationship between the local C and H compositions is addressed in Figure 7b. The average H concentration for all voxels of a given C concentration in the range of 0 to 12 % in steps of 2 % is plotted. The shaded error bar represents the standard error with respect to the mean in each case. A clear linear upward trend is observed ($X_{H(ave)} = 0.017\ X_C + 0.12$), confirming that the H composition increases, on average, with increasing C composition. This agrees with the partitioning behaviour of H observed in Figure 5 and adds a level of quantification to the change of H solubility with C content. This would indicate that the decomposed cementite, which contains a high density of crystalline defects, i.e. vacancies and dislocations, can indeed accommodate more H within its structure. In pristine cementite H solubility should be significantly lower.

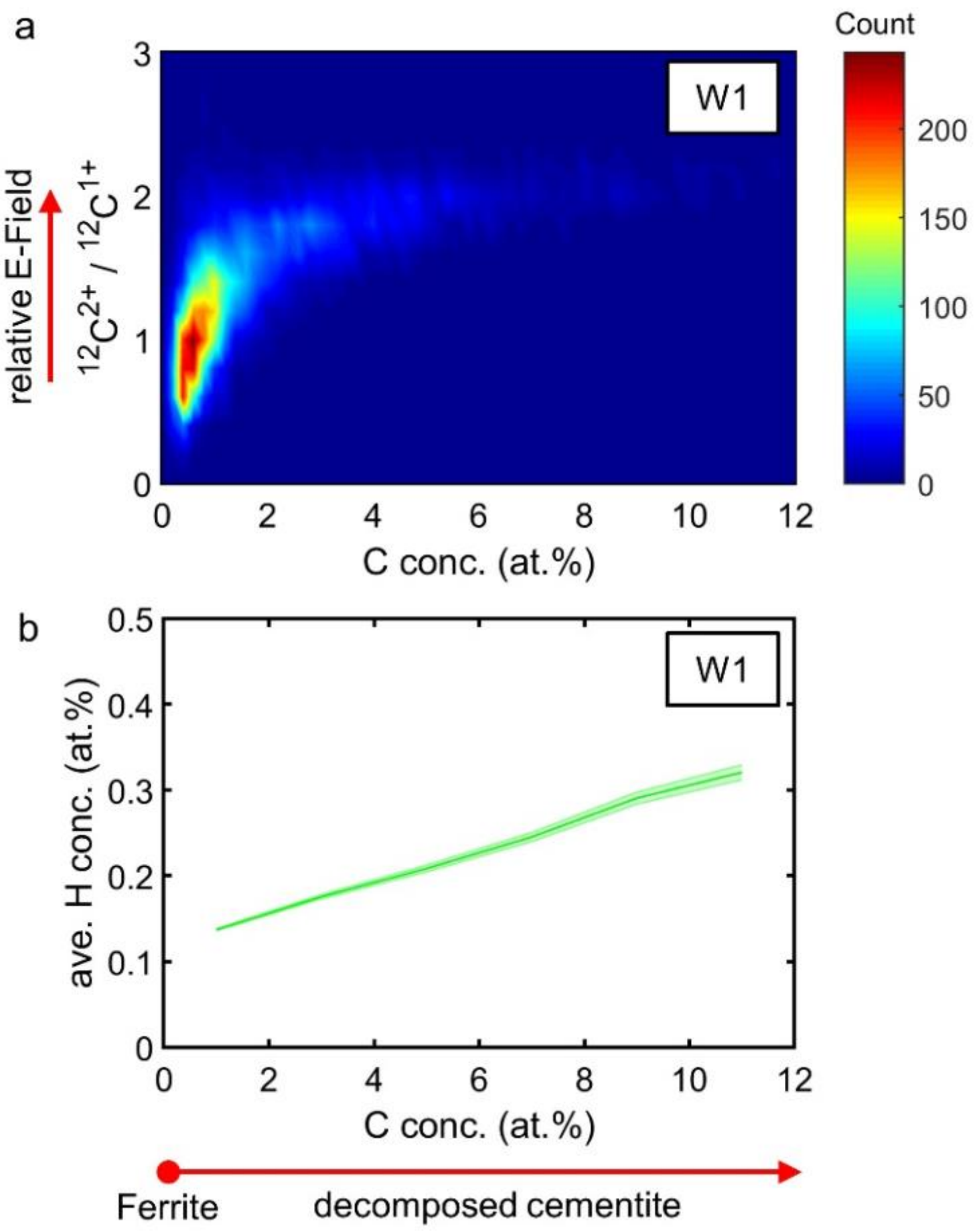

*Figure 7: (a) Relative C conc. (at. %) vs. relative E-field intensity histogram (b) C conc. (at. %) vs. average. H conc. (at. %) line graph with shaded standard error of the mean (SEM) error zone. Cuboidal voxel sizes of $8^3$ nm$^3$ and $2^3$ nm$^3$ were used for (a) and (b) respectively.*

## 3.8 Validating successful detection of D

In order to confirm the successful detection of deuterium in W4 and W5, the relative electric field and concentration of ions originating from the m/q = 2 Da peak were considered. Molecular ions are more likely to form at lower electric fields, so if the datasets corresponding to W4 and W5 showed regions of significantly lower field relative to the other datasets, this could be an indication that molecular hydrogen ($H_2$) rather than deuterium was being detected. The grey-scale in the background of Figure 8a is a 2D intensity histogram of all non-charged voltage-pulsed datasets (W1, W3, W6, W7, W1 + 3 (1), W1 + 3 (3)) with, superimposed, scatter plots of the two deuterium charged datasets (W4 and W5) in red and green. For all non-charged, voltage pulsed runs, the $^{12}C^{2+}/^{12}C^{1+}$ ratio is almost entirely between 0.4 – 2.3. The calculated $^{12}C^{2+}/^{12}C^{1+}$ ratios in each voxel for W4 and W5 also predominantly fall into this range. This is highlighted in the corresponding histogram for W4 and W5 displayed to the side, which again show no signs of $^{12}C^{2+}/^{12}C^{1+}$ ratios outside the range observed in the non-charged datasets. The average $^{12}C^{2+}/^{12}C^{1+}$ ratios were 1.2 and 1.5 in W4 and W5 respectively. No clear clusters of voxels with low field are observed. These observations point to similar field evaporation conditions and allows us to make direct comparisons between datasets. The scatter in W4 is similar to that observed in the non-charged voltage pulsed runs. The relative number of ions at m/q = 2 Da in workflow W5 is noticeably higher than that of the other voltage pulsed datasets with similar electric field, pointing to the importance of the cryogenic transfer to limit outgassing during specimen transfer.

Overall, the results suggest that a statistically significant number of ions originating from m/q = 2 Da were observed for the D-charged workflows at electric field levels very similar to that observed in the non-charged datasets suggesting that the peaks were likely indicating successful D detection. A region-of-interest approximately 1800 $nm^3$ around the cluster of m/q = 2 Da ions in W4 (the ‘local’ region which was indicated in Figure 5d) returned $^{12}C^{2+}/^{12}C^{1+}$ = 1.6 and a m/q = 2 Da concentration of 0.1 at. %. Figure 8b shows the reconstruction obtained from the W4 dataset as well as a cross sectional view of a thin slice through the tomogram showing the distribution of C and D atoms. In the corresponding compositional map, the local higher composition of D appears higher at one of the interfaces between the ferritic matrix and the decomposed cementite. Modelling work indicated a possible influence of the crystallographic character of the interface on the H/D distributions [36].

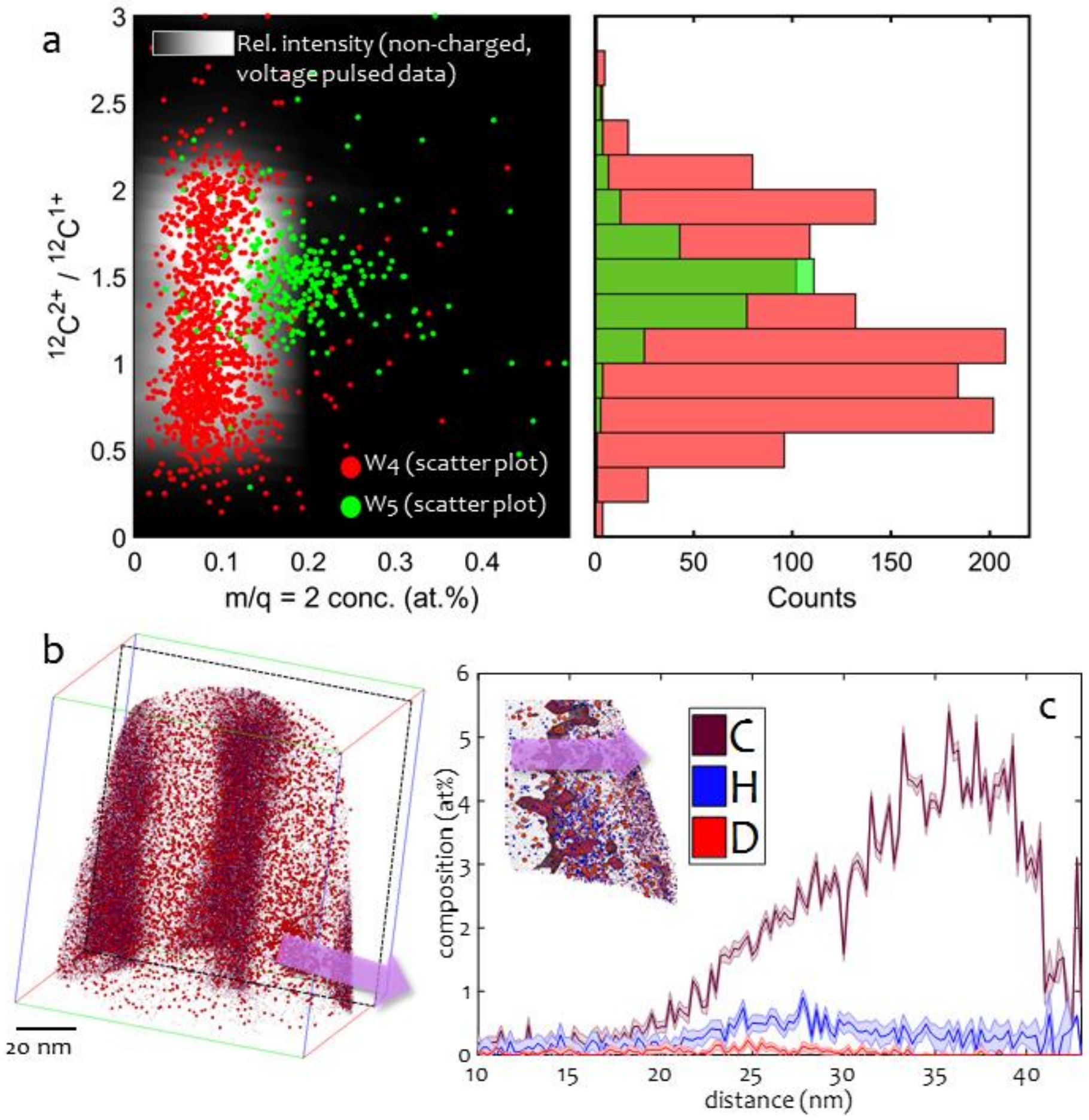

*Figure 8: (a) ion concentration at m/q = 2 vs. relative field ($^{12}C^{2+}$/ $^{12}C^{1+}$). A 2D histogram of all non-charged voltage pulsed runs (W1, W3, W6, W7, W1 +3 (1), W1 +3(3)) was calculated from the 8 nm voxelised reconstructions and the relative intensity plotted. Scatter plots for each $8^3$ $nm^3$ voxel in the D charged voltage pulsed runs (W4 and W5) were then superimposed for comparison). The corresponding histograms of the relative fields calculated for each of the $8^3$ $nm^3$ voxels in W4 and W5. (b) C, H and D distribution in the tomographic reconstruction obtained from the data obtained through W4. (c) composition profile calculated along the pink arrow, and, as shown in the inset, H and D tend to be more highly concentrated in the region with slightly less C within the decomposed cementite.*

## 3.9 Origins of the H signal detected in APT

The general perspective on studying H by APT was that it was impossible to distinguish H originating either from the specimen or from the chamber. While heating experiments suggested that most of the H being detected in the initial experiment (W1) was, in-fact, from the sample itself, it is still difficult to confirm exactly where this H was introduced. If exposure to atmosphere could lead to H ingress, likely to a small extent, H uptake could well occur during the specimen preparation. However, recent evidence has pointed to the introduction of significant amounts of H in Ti and Ti-alloys during FIB-based specimen preparation[18,22]. These reports agreed well with the literature that pointed to the formation of hydrides from mechanical and electrochemical polishing and ion beam-based preparation of specimens for microscopic investigations[37–39]. In steel and ferrous-alloys, no such reports exist, but there is no reason that H ingress would not occur.

The bare metallic surface initially produced by the dissolution of the metal or surface oxides offers many new, potentially very active, surface sites. Indeed, surface atoms are usually unsaturated and sterically more free to bond with other species. These atoms would also be the first to leave the structure during dissolution. The dissolution of iron in an acidic solution should in principle follow these steps

(1) An anodic reaction such as $Fe \rightarrow Fe^{2+} + 2\ e^{-}$or $Fe \rightarrow Fe^{3+} + 3\ e^{-}$ leading to the formation of ferric or ferrous ions in solution depending on e.g. the pH conditions;
(2) A cathodic reaction, likely involving the reduction of H and the release of gaseous $H_2$, i.e. $2H^{+} + 2e^{-} \rightarrow H_2(g)$

The electrochemical polishing process is usually accompanied by the formation of gaseous species, often assumed to be originating from the oxygen evolution reaction, leading to bubbling of $O_2$ at the dissolving anode.

However, this dissolution reaction will be accompanied by a vast array of possible intermediate steps reactions, including the formation of oxides or hydroxides, for instance:

$$Fe + H_2O \rightarrow FeO + 2\ H^{+} + 2e^{-}$$
$$2Fe + 3H_2O \rightarrow Fe_2O_3 + 6H^{+} + 6e^{-}$$

$$Fe + H_2O \rightarrow Fe(OH)^{-} + H^{+}$$
$$\text{or } Fe + 2H_2O \rightarrow \mathrm{H}Fe\mathrm{O}_2^{-} + 3\ H^{+} + 2e^{-}$$

In the case of the perchloric acid used here, the chlorine-containing ions can also react with the metal:

$$Fe + 2HClO_4 = Fe(ClO_4)_2 + H_2$$

$$2Fe + 6HClO_4 \rightarrow +3Fe(ClO_4)_3 + H_2$$

Under the conditions of very high potential used during the electrochemical polishing of the specimen, all of these reactions, likely only amongst others, can take place, albeit with different rates. No effort has ever been dedicated to controlling the conditions so as to select a specific reaction. In addition, literature points to the key role played by H in these reactions[40]. The presence of protons produced by these reactions at or near the surface can lead to the introduction of H into the material.

Importantly, both the adsorption of water on the surface as well as the formation of surface oxide will lead to an over concentration of protons in the layer at the interface between the metal surface and the solution. Some of these protons will also end up subject to chemisorption or physisorption at the metal surface. This makes their absorption by the metal not only possible but also likely. H penetration was shown to be energetically favourable by ab-initio simulations in the case of an iron surface in a sour gas environment [41], in conditions that would not be too dissimilar to those encountered here. H-ingress is also made more likely by the overconcentration of $H^{+}$ in the interfacial layer, i.e. the system will try to equilibrate the chemical potential of H in between the solution and the metal, leading to the introduction of H within the metal. This agrees with recent modelling effort by ab initio simulations of the corrosion of Mg showed that it was possible that the hydrogen evolution reaction could take place also at the anode, but also that the splitting of water at the metal surface can lead to the introduction of H within the metal[42].

There could be specimen preparation strategies optimised to limit or avoid uncontrolled introduction of H by using either highly controlled conditions of potential during electrochemical polishing, i.e. maybe in the range of 1V. An alternative could be to find conditions for controlled polishing in proton-free electrolytes or maybe in ionic liquids, including those containing H strongly bonded, i.e. part of C-H bonds for instance.

## 3.10 Characterising and quantifying H by atom probe tomography

The wide array of preparation and analysis protocols presented and discussed herein enable us to identify adequate ways to probe H and D at the near-atomic scale and do so with clear measures of distinction between intruded, adsorbed and generically trapped H/D atoms. This leads us to define the pathway to finally overcoming the long-term pending challenge of how to measure single H and D atoms in complex microstructures and how to quantify the associated errors.

Traditionally, the H peak observed at m/q = 1 Da in the mass spectrum of APT has been attributed to H present in the analysis chamber which gets field evaporated in the presence of the high electric field at the apex of the atom probe needle-shaped specimen. However, on closer inspection of our experimental data from W3, it appears that at least 80% of the H detected is from the specimen itself, whether from within the core of the specimen or adsorbed at the surface. This observation agrees with recent studies on other alloy systems which have also suggested that the H being detected at m/q = 1 Da most likely originates from the specimen[21,22]. It follows then that the H distribution in the datasets before heating is statistically significant enough to highlight real microstructural segregation and partitioning within the alloy itself, albeit with no control over the conditions of H intake, i.e. the chemical potential that led to the ingress. The low H composition detected after heating (W3) can then be attributed to a small amount of H from the chamber being field ionized or field desorbed but also to H in deep traps[43–45], unable to escape at the relatively low temperature (150°C) used for the H-release treatment.

The conditions during specimen preparation are highly out-of-equilibrium, which makes the determination of the chemical potential of H difficult if not impossible. Such conditions are not representative of H-uptake during the service life of a part. In addition, the baseline level of H originating from residual gas from the chamber cannot be accurately determined or predicted, so D-charging is still particularly useful, not only because it mitigates overlap, but also because it enables a more controlled source of D. D-charging can likely only be performed reliably after electrochemical polishing if a H-release heat treatment (W3) is performed.

Yet, under experimental conditions where the electrostatic field is optimised to lead to satisfactory analysis conditions, it is possible that the distribution of H, in particular correlated to regions of high electrostatic field, reveals the actual partitioning behaviour of H within the material [21]. Our experimental results obtained via W2 however indicate that results obtained from laser pulsed APT cannot be trusted. The heating of the specimen subsequent to laser light absorption enables field evaporation at electric fields that are too low to shield the apex from H migrating along the specimen's shank[8], leading to excessive detection of $H_2^+$, some of which could dissociate into two $H^+$, making the characterisation of H doubtful and its reliable quantification impossible. The additional heat can also assist the surface diffusion of adsorbed atomic or molecular H at the surface towards the apex.

### 3.11 Rationalising pearlite's resistance against hydrogen embrittlement

With more stringent constraints on fuel efficiency for e.g. automotive vehicles, there has been a surge of interest in developing high-strength steels. However, these materials are critically subject to hydrogen embrittlement [46]. Most experimental studies of hydrogen embrittlement have focused on the mechanical degradation of the material in conjunction with indirect measurements of the effect of hydrogen on the mechanical behaviour of the alloy [6,47]. This was often supported by estimations of the content of hydrogen within a material, via for instance TDS, and made assumptions on the nature of the energy traps associated to specific microstructural features [4,47]. Amongst high-strength steels, as-drawn pearlitic steel wire have been reported to exhibit superior resistance against hydrogen embrittlement [6,48]. This was ascribed to the high density of H trapping sites in pearlitic steel associated to the high density of defects in the off-stoichiometric cementite [6], as well as possibly at interfaces between the ferrite and the cementite [49].

The results reported herein provide insights into the behaviour of H within the microstructure of these highly deformed pearlitic steels and the influence that the experimental workflow has on the final observed distribution. Only a relatively low number of hydrogen ions were detected in all datasets, even before any heating experiments, but the one obtained with laser pulsing (W2). This was expected due to the known low solubility of H in cementite and ferrite in steel [50]. The H and C atom maps, in e.g. Figure 5, indicate a clear partitioning behaviour of H to the decomposed cementite regions, and we find a clear trend between H and C content (Figure 7). We thus find that the hydrogen (or deuterium) atoms are primarily trapped within the cementite. Among the different possible mechanisms for this preference the cementite's dislocation content and C-vacancies are conceivable as trapping sites. D-charging after H-release supports the partitioning of D to the decomposed cementite, and a slight segregation is observed at an interface between the cementite and the ferritic matrix.

# 4 Summary and conclusion

To summarise, the distribution of H/D distribution in heavily cold-drawn pearlitic steel wire of hypereutectoid composition (Fe-0.98 C-0.31Mn-0.20Si-0.20Cr-0.01Cu0.006 P-0.007 S, wt.%) was studied using atom probe tomography (APT). Data from 7 different workflows were compared to elucidate hydrogen absorption and desorption behaviour in these materials and gain an understanding how various experimental decisions can significantly influence measured results.

The main findings can be summarized as follows:

- > 80 % reduction in H concentration was observed after a low temperature heat treatment in vacuum (150 °C for 1 hr) (W3, W1+3) suggesting that the majority of H being detected in the control dataset (W1) obtained via electropolishing and voltage pulsing was from within the sample itself and not gaseous H in the chamber.
- Laser pulsing (W2) should, in principle, be avoided as it results in enhanced ionization of residual gaseous H or desorption of residual H adsorbed on the cold surface of the specimen, both leading to a higher overall H level at 1 Da and 2 Da, inhibiting the detection of H and/or D from the specimen;
- Electrochemical polishing, as well as FIB-based preparation, result in the introduction of H within the microstructure, to a much lesser extent using cryogenic specimen preparation (W5);

- Electrochemical charging successfully introduced D into the microstructure, as confirmed by complementary TDS and observations of a peak at 2 Da after W4 and W5. The electric field conditions obtained for these workflows is not likely to cause the detection of molecular $H_2$ species; more successful charging is achieved after a low temperature annealing treatment to release H and free up potential trapping sites. In W5, which captured a decomposed cementite region the highest bulk D concentration of 0.75 at. % was observed.

To conclude, APT can be used to analyse the distribution of H within steels but the influence of sample preparation and transport as well as experimental running conditions must be taken into account. We can conclude from our observations that W4 is the most appropriate workflow for accurate analysis of H/D in pearlitic steel, and likely in other materials. Only under such optimised experimental conditions can the distribution of H from within the specimen be used to infer any microstructural information, in particular with regards to possible trapping sites, bearing in mind that the chemical potential of H during the electrochemical processing is likely unique to these conditions and not representative of conditions faced by the material in service. Ferrite and cementite were observed to both have a very low H solubility (bulk H concentrations in voltage pulsing runs, even after electrochemical charging attempts, were typically less than 0.1 at. %). H partitioned to the decomposed cementite regions, with some indication of a slight interfacial segregation, and the local concentration of H increased linearly with C concentration. Our observations rationalise the good resistance of severely deformed pearlitic steels to hydrogen-embrittlement.

## Contributions

A.B., B.G., D.R., M.H. designed the study. A.B. prepared specimens and performed APT. Y.L. and M.H. provided the pearlitic samples, helped with APT running conditions and performed TEM. L.T.S. supported with charging, cryo experiments including cryo-transfers and data interpretation. B.S. performed TDS analyses. L.T.S. and O.K. contributed to discussions on electrochemical polishing and charging. A.B. and B.G. processed the raw data, interpreted the APT results, wrote routines for data extraction, prepared figures. A.B., B.G. and D.R. drafted the manuscript. All authors discussed the results, had input and commented on the manuscript.

## Acknowledgements

AB and BS are grateful for funding from the AvH Stiftung. MH acknowledges financial support from the German Federal Ministery for Research BMBF through grant 03SF0535. BG and LTS acknowledge financial support from the ERC-CoG-SHINE-771602. The APT group at MPIE is grateful to the Max-Planck Gesellschaft and the BMBF for the funding of the Laplace Project. The authors thank Dr. H. Yarita, from Suzuki Metal Industry Co., Ltd., for providing the cold drawn specimens. Yanhong Chang, Uwe Tezins, Andreas Sturm and Christian Bross are thanked for technical support with cryogenic transfer experiments. Waldemar Krieger, Daniel Haley, Eason Chen are thanked for help advising on electrochemical Deuterium charging experiments and interpretation of the electrochemical reaction occurring during charging. David Mayweg is thanked for help with the multiwire puck design. Frédéric De Geuser is acknowledged for helping with the code for voxelization of the APT data.

## Supplementary Information

Experiment inventory (note: experiments starting with R5096 were ran on a Cameca LEAP 5000 XR instrument with reflectron while experiments starting with R5076 were ran on a Cameca LEAP 5000 XS with straight flight path, both at the Max Planck Institute für Eisenforschung).

| Workflow | RunID | Ions reconstructed | Temp (K) | Pulse fraction (%) | Laser pulse energy (pJ) | Pulse rate (kHz) |
|---|---|---|---|---|---|---|
| 1 | R5096_33494 | 1.48e8 | 60 | 20 | N/A | 200 |
| 2 | R5076_33083 | 7.4e7 | 60 | N/A | 40 | 250 |
| 3 | R5096_35040 | 9.5e7 | 60 | 15 | N/A | 200 |
| 4 | R5096_33324 | 2.8e7 | 60 | 20 | N/A | 200 |
| 5 | R5096_35122 | 5.9e6 | 60 | 15 | N/A | 200 |
| 6 | R5096_37265 | 3.3e7 | 60 | 15 | N/A | 200 |
| 7 | R5096_33582 | 6.9e6 | 60 | 20 | N/A | 200 |
| 1+3 (1) | R5096_37605 | 5.7e6 | 60 | 15 | N/A | 200 |
| 1+3 (2) | R5096_37611 | 8.8e6 | 60 | 15 | N/A | 200 |

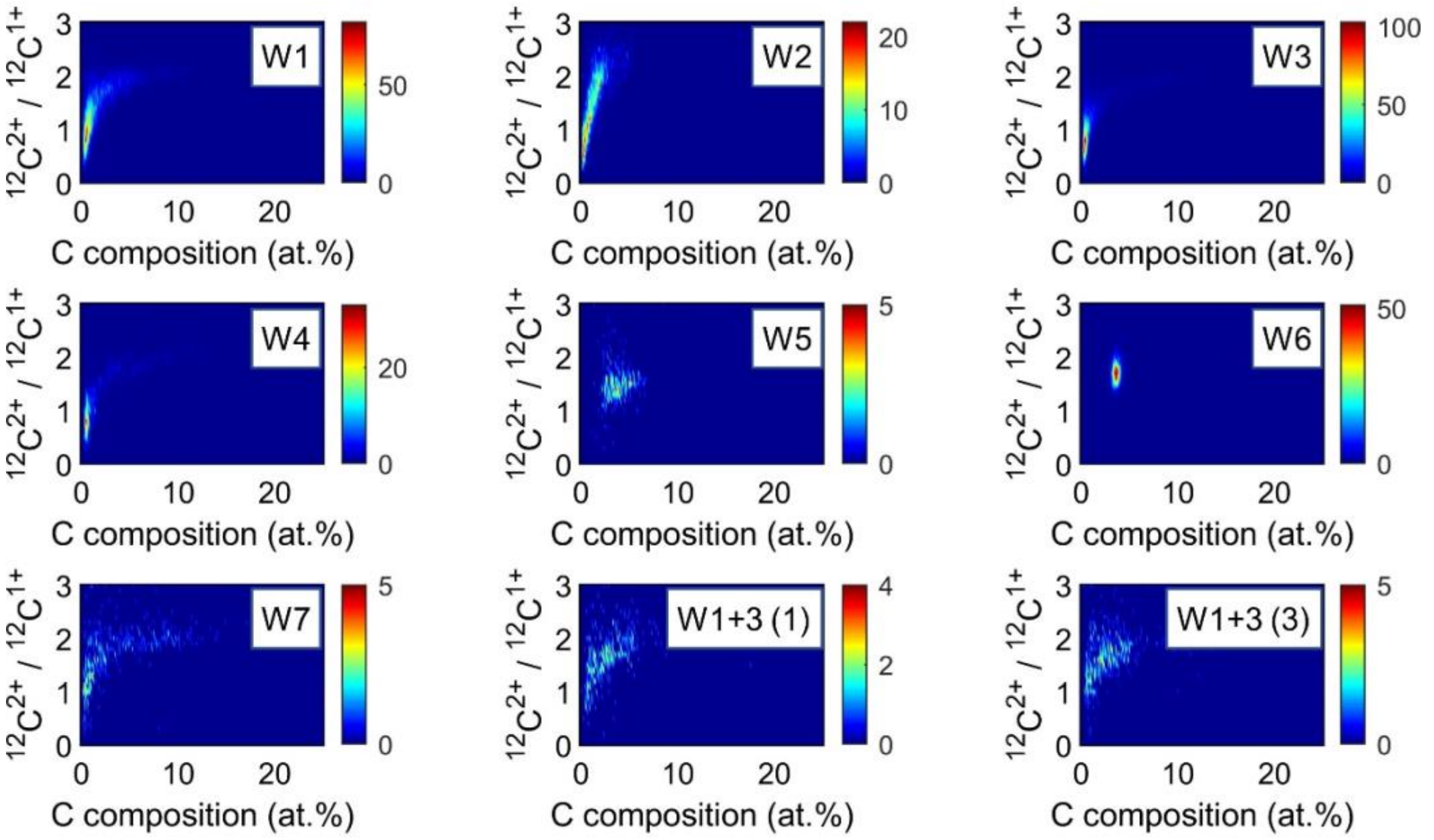

*Figure 9: C concentration vs. relative field ( $^{12}C^{2+}/^{12}C^{1+}$) for each of the datasets from the different workflows.*

**Mass spectrums**

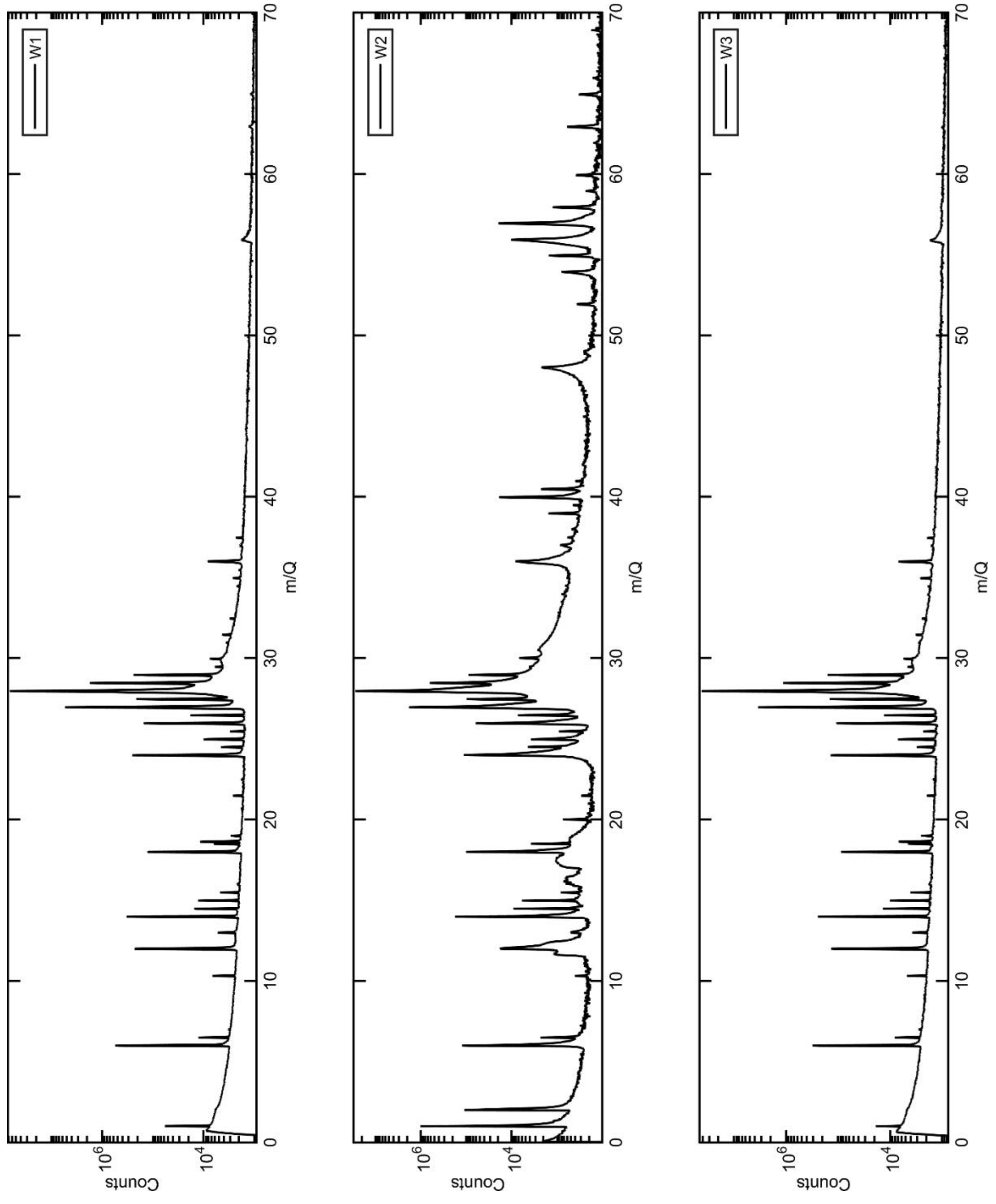

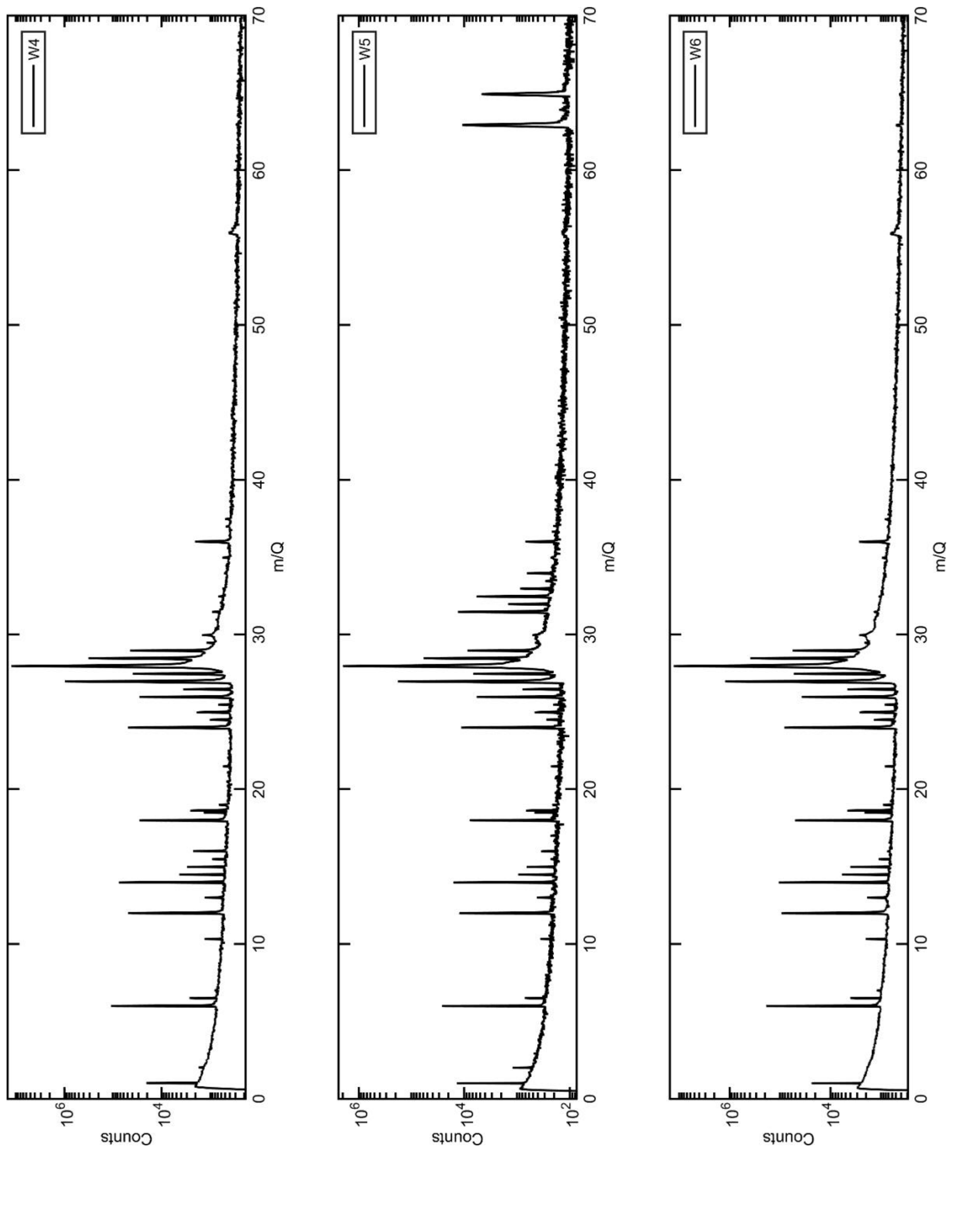
W4
W5
W6
Counts
m/Q

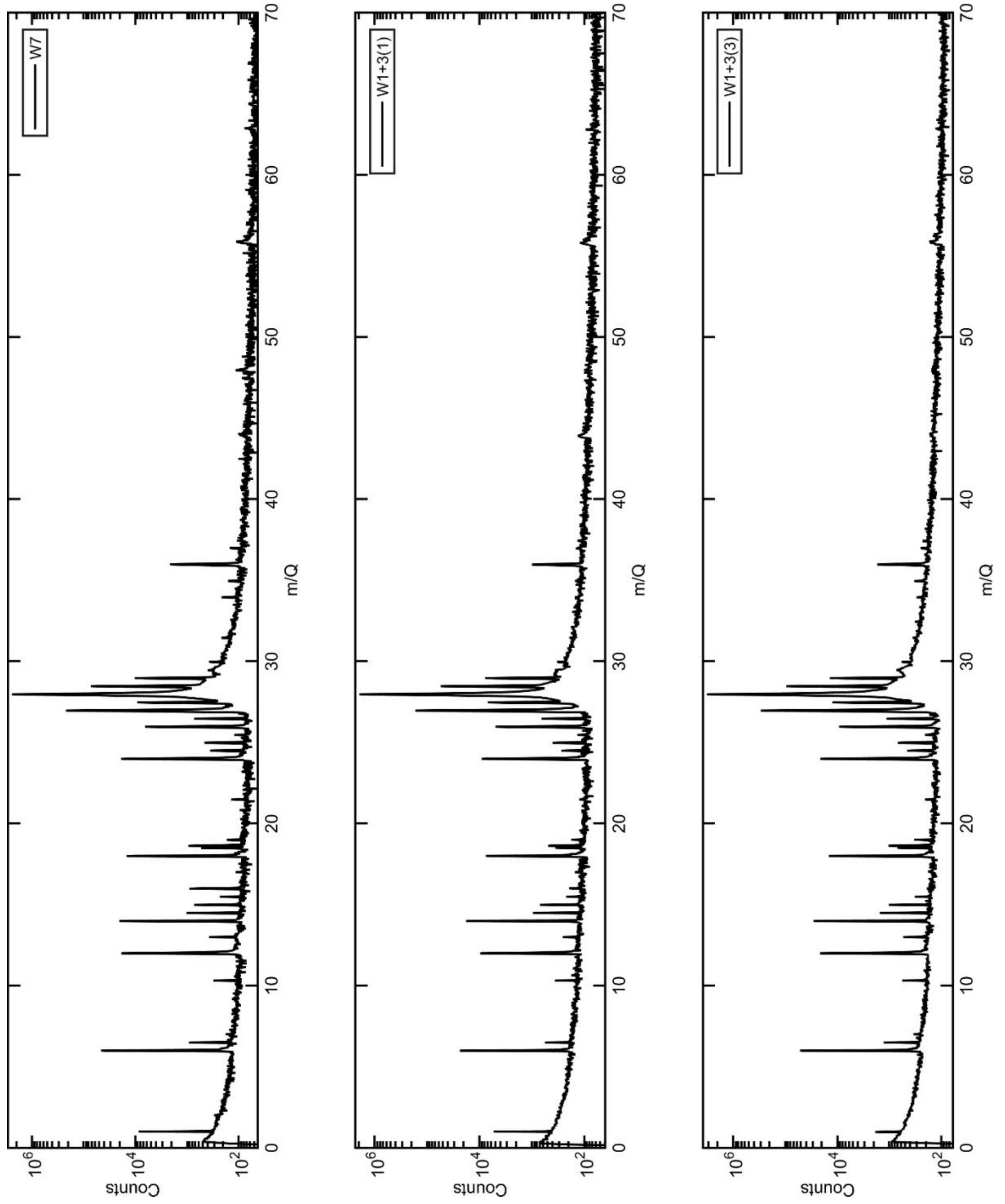